\documentstyle[proc]{rspublic}
\def\bi{\begin{itemize}}
\def\ei{\end{itemize}}
\def\bq{\begin{quotation}}
\def\eq{\end{quotation}}

\newcommand{\ket}[1]{\left | \, #1 \right \rangle}

\newcommand{\beq}{\begin{equation}}
\newcommand{\eeq}{\end{equation}}

\def\thedemobiblio#1{\smallskip\par
 \list{}{\labelwidth 0pt \leftmargin 1em \itemindent -1em \itemsep 1pt}
 \small \parindent 0pt
 \parskip 1.5pt plus .1pt\relax
 \def\newblock{\hskip .11em plus .33em minus .07em}
 \sloppy\clubpenalty4000\widowpenalty4000
 \sfcode`\.=1000\relax}

\begin{document}

\title[Quantum Algorithms Revisited]{Quantum Algorithms Revisited}

\author[R. Cleve, A. Ekert, C. Macchiavello and M. Mosca] {R.
  Cleve$^1$, A. Ekert$^2$, C. Macchiavello$^{2,3}$ and M. Mosca$^{2,4}$} 

\affiliation{ $^2$ Clarendon Laboratory, Department of Physics,
  University of Oxford,\\ Parks Road, Oxford OX1 3PU, U.K.\\ $^1$
  Department of Computer Science, University of Calgary\\ Calgary,
  Alberta, Canada T2N 1N4.\\ $^3$ I.S.I. Foundation, Villa Gualino, Viale Settimio
  Severo 65, 1033 Torino, Italy. \\$^4$ Mathematical Institute, University of Oxford, 24-29 St. Giles', Oxford OX1 3LB, U.K.}

\maketitle

\input{psfig.sty}

\begin{abstract}
  Quantum computers use the quantum interference of different
  computational paths to enhance correct outcomes and suppress
  erroneous outcomes of computations. A common pattern underpinning
  quantum algorithms can be identified when quantum computation is
  viewed as multi-particle interference. We use this approach to
  review (and improve) some of the existing quantum algorithms and to
  show how they are related to different instances of quantum phase
  estimation. We provide an explicit algorithm for generating any
  prescribed interference pattern with an arbitrary precision.
\end{abstract}

\section{Introduction}
Quantum computation is based on two quantum phenomena: quantum
interference and quantum entanglement. Entanglement allows one to
encode data into non-trivial multi-particle superpositions of some
preselected basis states, and quantum interference, which is a {\em
  dynamical process}, allows one to evolve initial quantum states
(inputs) into final states (outputs) modifying intermediate
multi-particle superpositions in some prescribed way. Multi-particle
quantum interference, unlike single particle interference, does not
have any classical analogue and can be viewed as an inherently quantum
process.

It is natural to think of quantum computations as multi-particle
processes (just as classical computations are processes involving
several ``particles'' or bits).  It turns out that viewing quantum
computation as multi-particle interferometry leads to a simple and
a unifying picture of known quantum algorithms. In this language
quantum computers are basically multi-particle interferometers with
phase shifts that result from operations of some quantum logic gates.
To illustrate this point, consider, for example, a Mach-Zehnder
interferometer (Fig. 1a).

A particle, say a photon, impinges on a half-silvered mirror, and,
with some probability amplitudes, propagates via two different paths
to another half-silvered mirror which directs the particle to one of
the two detectors.  Along each path between the two half-silvered
mirrors, is a phase shifter.  If the lower path is labelled as state
$\ket{0}$ and the upper one as state $\ket{1}$ then the state of the
particle in between the half-silvered mirrors and after passing
through the phase shifters is a superposition of the type
$\frac{1}{\sqrt 2} (\ket{0} + e^{i(\phi_1 - \phi_0)}\ket{1})$, where
$\phi_0$ and $\phi_1$ are the settings of the two phase shifters.
This is illustrated in Fig. 1a.  The phase shifters in the two paths
can be tuned to effect any prescribed relative phase shift $\phi =
\phi_1 - \phi_0$ and to direct the particle with probabilities
$\frac{1}{2} (1+ \cos\phi)$ and $\frac{1}{2} (1-\cos\phi)$
respectively to detectors ``0" and ``1".  The second half-silvered
mirror effectively erases all information about the path taken by the
particle (path $\ket{0}$ or path $\ket{1}$) which is essential for
observing quantum interference in the experiment.

\begin{figure}[!hb]
  \vspace*{0.2cm} \centerline{\psfig{width=8cm,file=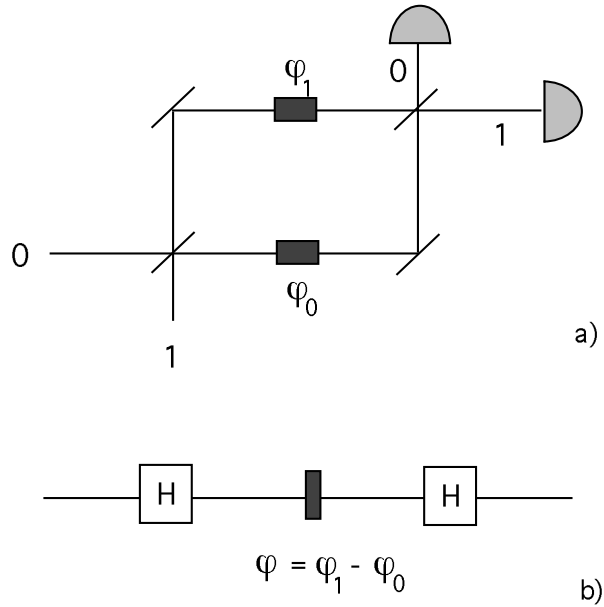}}
  \vspace*{5mm}
\caption{(a) Scheme of a Mach-Zehnder interferometer with two phase shifters. The
  interference pattern depends on the difference between the phase
  shifts in different arms of the interferometer. (b) The
  corresponding quantum network representation.}
\label{fig1}
\end{figure}

Let us now rephrase the experiment in terms of quantum logic gates.
We identify the half-silvered mirrors with the single qubit {\em Hadamard}
transform ($H$), defined as
\begin{eqnarray}
  \ket{0}&&\stackrel{H}{\longrightarrow} \textstyle{\frac{1}{\sqrt 2}}
(\ket{0} +\ket{1}) \nonumber \\
  \ket{1}&&\stackrel{H}{\longrightarrow} \textstyle{\frac{1}{\sqrt 2}}
(\ket{0} -\ket{1})\;.
\end{eqnarray}
The Hadamard transform is a special case of the more general 
{\em Fourier} transform, which we shall consider in Sect.~\ref{s:ft}.

We view the phase shifter as 
a single qubit gate. The resulting network corresponding to the
Mach-Zehnder interferometer is shown in Fig. 1b. The phase shift can
be ``computed'' with the help of an auxiliary qubit (or a set of
qubits) in a prescribed state $\ket{u}$ and some controlled-$U$
transformation where $U\ket{u}= e^{i\phi}\ket{u}$ (see Fig. 2).  Here
the controlled-$U$ means that the form of $U$ depends on the logical
value of the control qubit, for example we can apply the identity 
transformation to the auxiliary qubits (i.e. do nothing) when the
control qubit is in state $\ket{0}$ and apply a prescribed $U$ when
the control qubit is in state $\ket{1}$. The controlled-$U$ operation
must be followed by a transformation which brings all computational
paths together, like the second half-silvered mirror in the
Mach-Zehnder interferometer. This last step is essential to enable the
interference of different computational paths to occur---for example,
by applying a Hadamard transform.  In our example, we can
obtain the following sequence of transformations on the two qubits

\begin{eqnarray}
  \ket{0}\ket{u} \stackrel{H}{\longrightarrow} 
  \textstyle{\frac{1}{\sqrt 2}}(\ket{0} + \ket{1})\ket{u} & 
  \stackrel{c-U}{\longrightarrow} &
  \textstyle{\frac{1}{\sqrt 2}}(\ket{0} + e^{i\phi}\ket{1}) 
  \ket{u}\nonumber\\ &
  \stackrel{H}{\longrightarrow}& (\cos\textstyle{\phi \over 2}\ket{0} 
 - i \sin\textstyle{\phi \over 2}\ket{1})e^{i {\phi \over 2}} \ket{u}.
\label{sequ}
\end{eqnarray}


\begin{figure}[!hb]
  \vspace*{0.2cm} \centerline{\psfig{width=8cm,file=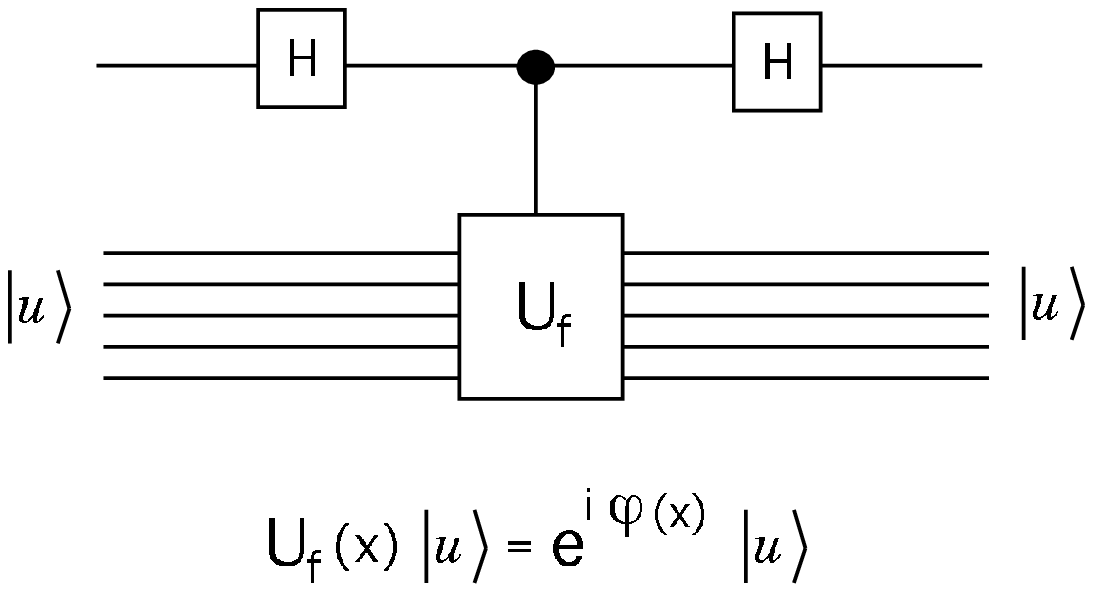}}
  \vspace*{5mm}
\caption{Network representation for the phase shift transformation of 
  Eq.~(\protect\ref{sequ}).  Here $x$ is a label for the state of the
  first qubit.}
\label{fig2}
\end{figure}

We note that the state of the auxiliary register $\ket{u}$, being an
eigenstate of $U$, is not altered along this network, but its
eigenvalue $e^{i\phi}$ is ``kicked back'' in front of the $\ket{1}$
component in the first qubit. The sequence (\ref{sequ}) is the exact
simulation of the Mach-Zehnder interferometer and, as we will illustrate in
the following sections, the kernel of quantum algorithms.

The rest of the paper is organised as follows. In the next section we
discuss Deutsch's problem (1985) which shows how differentiation between 
interference patterns (different phase-shifts) can lead to the formulation 
of computational problems.
Then, in Sect.~\ref{s:gda}, we review, in a unified way, generalisations of
Deutsch's problem, and propose further ones.
In Sect.~\ref{s:ft} we discuss an alternative and convenient way to 
view the quantum Fourier transform.
In Sect.~\ref{s:ip} we propose an efficient method for phase estimation based
on the quantum Fourier transform.
In order to illustrate how some of the existing algorithms can be
reformulated in terms of the multi-particle interferometry and the
phase estimation problem, in Sect.~\ref{s:shor} we rephrase Shor's
order-finding algorithm (used to factor) using the phase estimation
approach. 
Finally, in Sect. \ref{s:gen} we present a
universal construction which generates any desired interference
pattern with arbitrary accuracy. 
We summarise the conclusions in Sect.~\ref{s:conc}.

\section{Deutsch's Problem}
\label{s:da}

Since quantum phases in the interferometers can be introduced by some 
controlled-$U$ operations, it is natural to ask whether effecting 
these operations can be described as an interesting computational 
problem.
In this section, we illustrate how interference patterns 
lead to computational problems that are well-suited to quantum 
computations, by presenting the first such problem that was 
proposed by David Deutsch (1985).

To begin with, suppose that the phase shifter in
the Mach-Zehnder interferometer is set either to $\phi = 0$ or to
$\phi=\pi$.  Can we tell the difference? Of course we can. In fact, a
single instance of the experiment determines the difference: for
$\phi=0$ the particle {\em always} ends up in the detector ``0" and
for $\phi=\pi $ {\em always} in the detector ``1". Deutsch's 
problem is related to this effect.

Consider the Boolean functions $f$ that map $\{0,1\}$ to $\{0,1\}$.
There are exactly four such functions: two constant functions
($f(0)=f(1)=0$ and $f(0)=f(1)=1$) and two ``balanced'' functions
($f(0)=0, f(1)=1$ and $f(0)=1, f(1)=0$).
Informally, in Deutsch's problem, one is allowed to evaluate the 
function $f$ {\em only once} and required to deduce from the result 
whether $f$ is constant or balanced (in other words, whether the binary 
numbers $f(0)$ and $f(1)$ are the same or different).
Note that we are not asked for the particular values $f(0)$ and $f(1)$ 
but for a global property of $f$.
Classical intuition tells us that to determine this global property 
of $f$, we have to evaluate both $f(0)$ and $f(1)$ anyway, which 
involves evaluating $f$ twice.
We shall see that this is not so in the setting of quantum information, 
where we can solve Deutsch's problem with a single function evaluation, by 
employing an algorithm that has the same mathematical structure 
as the Mach-Zehnder interferometer.

Let us formally define the operation of ``evaluating'' $f$ in terms of 
the {\em $f$-controlled-NOT} operation on two bits: the first contains 
the input value and the second contains the output value.
If the second bit is initialised to $0$, the $f$-controlled-NOT maps 
$(x,0)$ to $(x,f(x))$.
This is clearly just a formalization of the operation of computing $f$.
In order to make the operation reversible, the mapping is defined for 
{\em all\/} initial settings of the two bits, taking $(x,y)$ to 
$(x,y \oplus f(x))$.
Note that this operation is similar to the controlled-NOT (see, 
for example, Barenco {\em et al.} (1995)), except that the second bit 
is negated when $f(x)=1$, rather than when $x=1$.

\begin{figure}[!hb]
  \vspace*{0.2cm} \centerline{\psfig{width=9cm,file=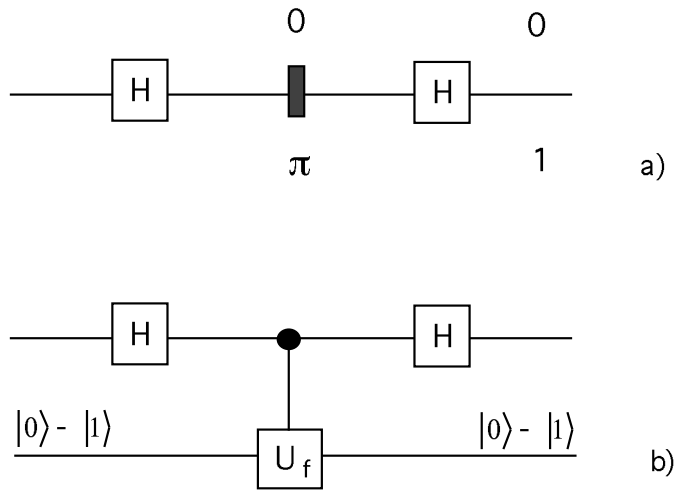}}
  \vspace*{5mm}
\caption{Network representation of Deutsch's algorithm.}
\label{deutsch}
\end{figure}

If one is only allowed to perform classically 
the $f$-controlled-NOT operation once, 
on any input from $\{0,1\}^2$, then it is {\em impossible} to distinguish 
between balanced and constant functions in the following sense.
Whatever the outcome, both possibilities (balanced and constant) 
remain for $f$.
However, if quantum mechanical superpositions are allowed then a 
single evaluation of the $f$-controlled-NOT suffices to classify $f$.
Our quantum algorithm that accomplishes this is best represented as 
the quantum network shown in Fig.~\ref{deutsch}b, 
where the middle operation is the $f$-controlled-NOT, whose semantics in 
quantum mechanical notation are
\begin{equation}
  \ket{x}\ket{y} \stackrel{f-c-N}{\longrightarrow} \ket{x}\ket{y
    \oplus f(x)}\;.
\label{eqf}
\end{equation}

The initial state of the qubits in the quantum network is
$\ket{0}(\ket{0}-\ket{1})$ (apart from a normalization factor, which
will be omitted in the following).  After the first Hadamard 
transform, the state of the two qubits has the form
$(\ket{0}+\ket{1})(\ket{0}-\ket{1})$.  To determine the effect of the
$f$-controlled-NOT on this state, first note that, for each $x \in
\{0,1\}$,
\begin{equation}
  \ket{x}(\ket{0}-\ket{1}) \stackrel{f-c-N}{\longrightarrow}
  \ket{x}(\ket{0\oplus f(x)}-\ket{1\oplus f(x)})=(-1)^{f(x)}
  \ket{x}(\ket{0}-\ket{1}) \;.
\end{equation}
Therefore, the state after the $f$-controlled-NOT is
\begin{equation}
  ((-1)^{f(0)}\ket{0}+(-1)^{f(1)}\ket{1}) (\ket{0}-\ket{1})\;.
\label{st-d}
\end{equation}
That is, for each $x$, the $\ket{x}$ term acquires a phase factor of $(-1)^{f(x)}$, 
which corresponds to the eigenvalue of the state of the auxiliary qubit under 
the action of the operator that sends $\ket{y}$ to $\ket{y \oplus f(x)}$.

This state can also be written as 
\begin{equation}
(-1)^{f(0)}(\ket{0}+(-1)^{f(0)\oplus f(1)}\ket{1})\;,
\end{equation}
which, after applying the second Hadamard transform, becomes
\begin{equation}
(-1)^{f(0)}\ket{f(0) \oplus f(1)}\;.
\end{equation}
Therefore, the first qubit is finally in state $\ket{0}$ if the 
function $f$ is constant and in state $\ket{1}$ if the function is 
balanced, and a measurement of this qubit distinguishes these cases 
with certainty.

This algorithm is an improved version of the first quantum algorithm 
for this problem proposed by Deutsch (1985), which accomplishes the 
following.
There are three possible outcomes: ``balanced'', ``constant'', and 
``inconclusive''.
For any $f$, the algorithm has the property that: with probability 
$1 \over 2$, it outputs ``balanced'' or ``constant'' (correctly 
corresponding to $f$); and, with probability $1 \over 2$, it outputs 
``inconclusive'' (in which case no information is determined about $f$).
This is a task that no classical computation can accomplish (with a single 
evaluation of the $f$-controlled-NOT gate).
In comparison, our algorithm can be described as {\em always} producing 
the output ``balanced'' or ``constant'' (correctly).
Alain Tapp (1997) independently discovered an algorithm for Deutsch's 
problem that is similar to ours.

Deutsch's result laid the foundation for the new field of quantum 
computation, and was followed by several other quantum algorithms 
for various problems, which all seem to rest on the same generic 
sequence: a Fourier transform, followed by an $f$-controlled-$U$, 
followed by another Fourier transform.
(In some cases, such as Lov Grover's ``database search'' 
algorithm (1996), this sequence is a critical component to a larger 
algorithm; see Appendix B).
We illustrate this point by reviewing several of these other algorithms 
in the sections that follow.

\section{Generalisations of Deutsch's Problem}
\label{s:gda}

Deutsch's original problem was subsequently generalised by Deutsch and 
Jozsa (1992) for Boolean functions $f : \{0,1\}^n\rightarrow
\{0,1\}$ in the following way.
Assume that, for one of these functions, it is ``promised'' 
that it is either constant or balanced (i.e. has an equal number of 
0's outputs as 1's), and consider the goal of determining which of the 
two properties the function actually has.

\begin{figure}[!hb]
  \vspace*{0.2cm} \centerline{\psfig{width=8.5cm,file=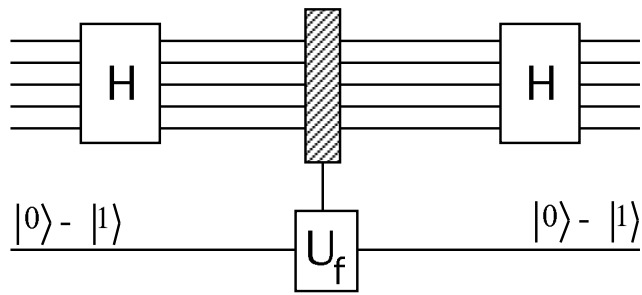}}
  \vspace*{5mm}
\caption{Network representation of Deutsch-Jozsa's and Bernstein-Vazirani's
  algorithms.}
\label{fig4}
\end{figure}

How many evaluations of $f$
are required to do this?  Any classical algorithm for this problem
would, in the worst-case, require $2^{n-1}+1$ evaluations of $f$
before determining the answer with certainty.  There is a quantum
algorithm that solves this problem with a single evaluation of $f$.
The algorithm is presented in Fig.~\ref{fig4}, 
where the control register is now composed of $n$ qubits, all initially in
state $\ket{0}$, denoted as $\ket{00\cdots 0}$, and, 
as in the quantum algorithm for Deutsch's simple problem, 
an auxiliary qubit is employed, which is initially set to state 
$\ket{0}-\ket{1}$ and is not altered during the computation.
Also, the $n$-qubit Hadamard transform $H$ is defined as 
\begin{equation}
  \ket{x} \stackrel{H}{\longrightarrow} \sum_{y \in
    \{0,1\}^n}(-1)^{x\cdot y}\ket{y}\;,
\end{equation}
for all $x \in \{0,1\}^n$, where 
\begin{equation}
x\cdot y = (x_1 \wedge y_1) \oplus \cdots \oplus (x_n \wedge y_n)
\end{equation}
(i.e.\ the scalar product modulo two).
This is equivalent to performing a one-qubit Hadamard transform 
on each of the $n$ qubits individually.
The actual computation of the function $f$ is by means of
an $f$-controlled-NOT gate (the middle gate in Fig.~\ref{fig4}), 
which acts as
\begin{equation}
\ket{x}\ket{y} \stackrel{f-c-N}{\longrightarrow} 
\ket{x}\ket{y \oplus f(x)}\;.
\end{equation}
This is similar to Eq.~(\ref{eqf}), except that now $x \in \{0,1\}^n$.

Stepping through the execution of the network, the state after the 
first $n$-qubit Hadamard transform is applied is 
\begin{equation}
  \sum_{x \in \{0,1\}^n}\ket{x}(\ket{0}-\ket{1})\;,
\end{equation}
which, after the $f$-controlled-NOT gate, is 
\begin{equation}
  \sum_{x \in \{0,1\}^n}(-1)^{f(x)}\ket{x}(\ket{0}-\ket{1})\;.
\label{dj2}
\end{equation}
Finally, after the last Hadamard transform, the state is 
\begin{equation}
  \sum_{x,y \in \{0,1\}^n}(-1)^{f(x) \oplus (x \cdot y)}\ket{y}
  (\ket{0}-\ket{1})\;.
\end{equation}

Note that the amplitude of $\ket{00\cdots 0}$ is
$\sum_{x\in \{0,1\}^n} \frac{(-1)^{f(x)}}{2^n}$
so if $f$ is constant then this state is $(-1)^{f(00\cdots\,
  0)}\ket{00\cdots 0}(\ket{0}-\ket{1})$; whereas, if $f$ is balanced
then, for the state of the first $n$ qubits, the amplitude of
$\ket{00\cdots 0}$ is zero.  Therefore, by measuring the first $n$
qubits, it can be determined with certainty whether $f$ is constant or
balanced.
Note that, as in Deutsch's simple example, this entails a single 
$f$-controlled-NOT operation.
(This is a slight improvement of Deutsch and Jozsa's original 
algorithm, which involves two $f$-controlled-NOT operations.)

Following Deutsch and Jozsa, Ethan Bernstein and Umesh Vazirani (1993) 
formulated a variation of the above problem that can be solved with 
the same network.  Suppose that $f : \{0,1\}^n \rightarrow \{0,1\}$ is 
of the form 
\begin{equation}
f(x) = (a_1 \wedge x_1) \oplus \cdots \oplus (a_n \wedge x_n) \oplus b 
= (a \cdot x) \oplus b\;,
\end{equation}
where $a \in \{0,1\}^n$ and $b \in \{0,1\}$, and consider the goal 
of determining $a$.
Note that such a function is constant if $a = 00 \cdots 0$ and 
balanced otherwise (though a balanced function need not be of this form).
Furthermore, the classical determination of $a$ requires at least $n$ 
$f$-controlled-NOT operations (since $a$ contains $n$ bits of 
information and each classical evaluation of $f$ yields a single bit 
of information).
Nevertheless, by running the quantum network given in Fig.~\ref{fig4}, 
it is possible to determine $a$ with a single $f$-controlled-NOT 
operation.

The initial conditions are the same as above.  In this case,
Eq.~(\ref{dj2}) takes the simple form
\begin{equation}
\sum_{x \in \{0,1\}^n}(-1)^{(a \cdot x) \oplus b}\ket{x}(\ket{0}-\ket{1})\;,
\label{dj4}
\end{equation}
which, after the final Hadamard transform, becomes
\begin{equation}
  (-1)^{b}\sum_{x,y \in
    \{0,1\}^n}(-1)^{x\cdot(a \oplus y)}\ket{y}(\ket{0}-\ket{1})\;,
\end{equation}
which is equivalent to $(-1)^{b}\ket{a} (\ket{0}-\ket{1})$.
Thus, a measurement of the control register yields the value of $a$.
(Bernstein and Vazirani's algorithm is similar to the above, 
except that it employs two $f$-controlled-NOT operations instead of one.
Also, this problem, and its solution, is very similar to the search 
problems considered by Barbara Terhal and John Smolin (1997).)

The network construction presented in this section (Fig.~\ref{fig4}) 
can be generalised to the case of a Boolean function 
$f : \{0,1\}^n\rightarrow \{0,1\}^m$ (with $m\leq n$), 
with the promise that the parity of the elements in the range of
$f$ is either constant or evenly balanced (i.e. its output values all
have the same parity, or half of them have parity $0$ and half have
parity $1$).  In this case, by choosing an auxiliary register composed
of $m$ qubits, and setting all of them in the initial state
$(\ket{0}-\ket{1})$, it is possible to solve the problem with
certainty in one run of the network.  As in the above case, the
function is constant when the $n$ qubits of the first register are 
detected in state $\ket{00 \cdots 0}$, and evenly balanced otherwise.

A particular subclass of the above functions consists of those that 
are of the form $f(x) = (A \cdot x) \oplus b$, where $A$ is an 
$m \times n$ binary matrix,  $b$ is a binary $m$-tuple, and $\oplus$ 
is applied bitwise (this can be thought of as an affine linear function 
in modulo-two arithmetic).
The output string of $f$ has constant parity if 
$(11 \cdots 1) \cdot A = (00 \cdots 0)$ and has balanced parity otherwise.
It is possible to determine all the entries of $A$ by evaluating the 
function $f$ only $m$ times, via a suitable multi-qubit $f$-controlled-NOT 
gate of the form 
\begin{equation}
\ket{x}\ket{y} \stackrel{f-c-N}{\longrightarrow} 
\ket{x}\ket{y \oplus f(x)}\;,
\end{equation}
where $x \in \{0,1\}^n$ and $y \in \{0,1\}^m$.
The network described below is a generalisation of that in Fig.~\ref{fig4}, 
and determines the $n$-tuple $c \cdot A$, where $c$ is any binary $m$-tuple.
The auxiliary register is composed of $m$ qubits, which are initialised 
to the state 
\begin{equation}
(\ket{0}+(-1)^{c_1}\ket{1})(\ket{0}+(-1)^{c_2}\ket{1}) \cdots 
(\ket{0}+(-1)^{c_{m}}\ket{1})\;.
\end{equation}
(This state can be ``computed'' by first setting the auxiliary register 
to the state $\ket{c_1 c_2 \cdots c_m}$ and then applying a Hadamard 
transform to it.)
The $n$-qubit control register is initialised in state $\ket{00 \cdots 0}$, 
and then a Hadamard transform is applied to it.
Then the $f$-controlled-NOT operation is performed, and is followed by 
another Hadamard transform to the control register.
It is straightforward to show that the control register will then 
reside in the state $\ket{c \cdot A}$.
By running the network $m$ times with suitable choices for $c$, 
all the entries of $A$ can be determined.
Peter H\o yer (1997) independently solved a problem that is similar 
to the above, except that $f$ is an Abelian group homomorphism, 
rather than an affine linear function.

\section{Another Look at the Quantum Fourier Transform}
\label{s:ft}

The quantum Fourier transform (QFT) on the additive group of integers
modulo $2^m$ is the mapping
\begin{equation}
  \ket{a} \stackrel{F_{2^m}}{\longrightarrow} \sum_{y=0}^{2^m-1}
  e^{\frac{2\pi i a y}{2^m}} \ket{y}\;,
\label{ft1}
\end{equation}
where $a \in \{0,\ldots,2^m-1\}$ (Coppersmith 1994). Let $a$ be
represented in binary as $a_1 \ldots a_m \in \{0,1\}^m$, where $a =
2^{m-1} a_1 + 2^{m-2} a_2 + \cdots + 2^{1} a_{m-1} + 2^0 a_m$ (and
similarly for $y$).

It is interesting to note that the state (\ref{ft1}) is unentangled,
and can in fact be factorised as
\begin{equation}
  (\ket{0} + e^{2\pi i (0.a_m)}\ket{1}) (\ket{0} + e^{2\pi i
    (0.a_{m-1}a_m)}\ket{1}) \cdots (\ket{0} + e^{2\pi i (0.a_1a_2
    \ldots a_m)}\ket{1})\;.
\label{ft2}
\end{equation}
This follows from the fact that
\begin{eqnarray}
  \lefteqn{e^{2\pi i a y \over 2^m}\ket{y_1 \cdots y_m}} & & \\ & = &
  e^{2\pi i (0.a_m) y_1}\ket{y_1} e^{2\pi i (0.a_{m-1}a_m)
    y_2}\ket{y_2} \cdots e^{2\pi i (0.a_1a_2 \ldots a_m)
    y_m}\ket{y_m}\;,
\end{eqnarray}
so the coefficient of $\ket{y_1 y_2 \cdots y_m}$ in (\ref{ft1})
matches that in (\ref{ft2}).

\begin{figure}[!hb]
  \vspace*{0.2cm} \centerline{\psfig{width=12cm,file=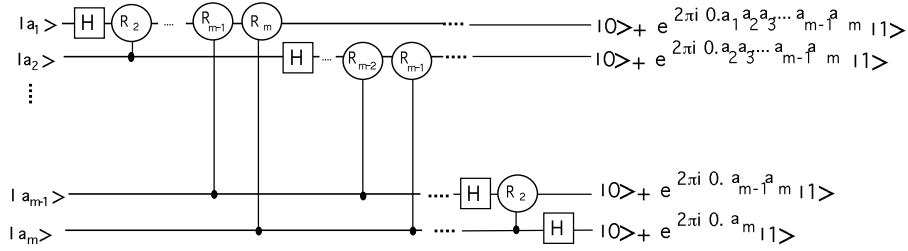}}
  \vspace*{0.06cm}
  \caption{A network for $F_{2^m}$ shown acting on the basis state 
  $\ket{a_1a_2 \cdots a_m}$. At the end, the order of the output qubits 
  is reversed (not shown in diagram).}
\label{fig5}
\end{figure}

A network for computing $F_{2^n}$ is shown in Fig.~\ref{fig5}.

In the above network, $R_k$ denotes the unitary transformation
\begin{equation}
  R_k = \pmatrix{ 1 & 0 \cr 0 & e^{2 \pi i / 2^k}}\;.
\end{equation}

We now show that the network shown in~Fig.~\ref{fig5} produces the
state (\ref{ft1}).  The initial state is $\ket{a} = \ket{a_1 a_2
  \cdots a_m}$ (and $a/2^m = 0.a_1a_2 \ldots a_m$ in binary).
Applying $H$ to the first qubit in $\ket{a_1 \cdots a_m}$ produces
the state
\[ (\ket{0} + e^{2\pi i (0.a_1)}\ket{1})\ket{a_2 \cdots a_m} .\] Then
applying the controlled-$R_2$ changes the state to
\[ (\ket{0} + e^{2\pi i (0.a_1a_2)}\ket{1})\ket{a_2 \cdots a_m} .\]
Next, the controlled-$R_3$ produces 
\[ (\ket{0} + e^{2\pi i (0.a_1a_2a_3)}\ket{1})\ket{a_2 \cdots a_m}, \]
and so on, until the state is 
\[ (\ket{0} + e^{2\pi i (0.a_1 \ldots a_m)}\ket{1})\ket{a_2 \cdots a_m} .\]
The next $H$ yields 
\[ (\ket{0} + e^{2\pi i (0.a_1 \ldots a_m)}\ket{1})
(\ket{0} + e^{2\pi i (0.a_2)}\ket{1})\ket{a_3 \cdots a_m} \]
and the controlled-$R_2$ to -$R_{m-1}$ yield
\begin{equation}
(\ket{0} + e^{2\pi i (0.a_1\ldots a_m )}\ket{1})
(\ket{0} + e^{2\pi i (0.a_2\ldots a_m)}\ket{1})\ket{a_3 \cdots a_m}\;.
\end{equation}
Continuing in this manner, the state eventually becomes 
\[ (\ket{0} + e^{2\pi i (0.a_1 \ldots a_m)}\ket{1}) (\ket{0} + e^{2\pi
  i (0.a_2 \ldots a_m)}\ket{1}) \cdots (\ket{0} + e^{2\pi i (0.a_m)
  }\ket{1})\;,\] 
which, when the order of the qubits is reversed, is state (\ref{ft2}).

Note that, if we do not know $a_1 \cdots a_m$, but are given a state
of the form (\ref{ft2}), then $a_1 \cdots a_m$ can be easily extracted
by applying the inverse of the QFT to the state, which will yield the
state $\ket{a_1 \cdots a_m}$.

\section{A Scenario for Estimating Arbitrary Phases}
\label{s:ip}

In Sect.~1, we noted that differences in phase shifts by $\pi$ can, in
principle, be detected exactly by interferometry, and by quantum
computations.  In Sects. 2 and 3, we reviewed powerful computational
tasks that can be performed by quantum computers, based on the mathematical 
structure of detecting these phase differences.
In this section, we consider the case of {\em arbitrary} phase differences, 
and show in simple terms how to obtain good estimators for them, 
via the quantum Fourier transform.
This phase estimation plays a central role in the fast quantum 
algorithms for factoring and for finding discrete logarithms 
discovered by Peter Shor (1994).
This point has been nicely emphasised by the quantum algorithms 
presented by Alexi Kitaev (1995) for the Abelian stabiliser problem.

Suppose that $U$ is any unitary transformation on $n$ qubits and 
$\ket{\psi}$ is an eigenvector of $U$ with eigenvalue $e^{2 \pi i \phi}$, 
where $0 \le \phi < 1$.
Consider the following scenario.
We do not explicitly know $U$ or $\ket{\psi}$ or $e^{2 \pi i \phi}$, 
but instead are given devices that perform controlled-$U$, 
controlled-$U^{2^1}$, controlled-$U^{2^2}$ (and so on) operations.
Also, assume that we are given a single preparation of the state 
$\ket{\psi}$. From this, our goal is to obtain an $m$-bit estimator of $\phi$.

\begin{figure}[!hb]
 \vspace*{0.4cm} \centerline{\psfig{width=12cm,file=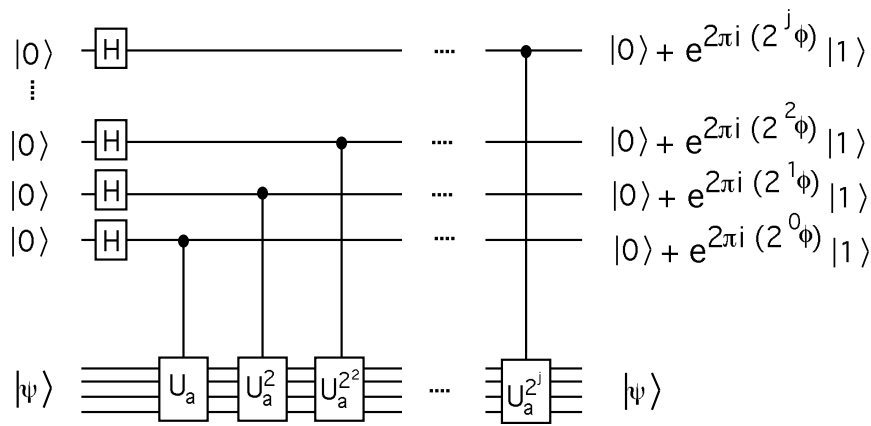}}
  \vspace*{1mm}
\caption{ A network illustrating estimation of phase $\phi$ with
  $j$-bit precision. The same network forms the kernel of the
  order-finding algorithm discussed in
  Section~\protect{\ref{s:shor}}.}
\label{shorkit}
\end{figure}

This can be solved as follows.
First, apply the network of Fig.~\ref{shorkit}.
This network produces the state
\begin{equation}
(\ket{0} + e^{2\pi i 2^{m-1} \phi}\ket{1})
(\ket{0} + e^{2\pi i 2^{m-2} \phi}\ket{1})
\cdots
(\ket{0} + e^{2\pi i \phi}\ket{1})  =  
\sum_{y=0}^{2^m-1} e^{2\pi i \phi y} \ket{y}\;.
\label{qftphi}
\end{equation}
As noted in the last section, in the special case where $\phi = 0.a_1
\ldots a_m$, the state $\ket{a_1\cdots a_m}$ (and hence $\phi$) can be
obtained by just applying the inverse of the QFT (which is the network 
of Fig.~\ref{fig5} in the backwards direction).
This will produce the state $\ket{a_1\cdots a_m}$ exactly 
(and hence $\phi$).

However, $\phi$ is not in general a fraction of a power of two (and
may not even be a rational number).
For such a $\phi$, it turns out that applying the
inverse of the QFT produces the best $m$-bit approximation of $\phi$
with probability at least $4 / \pi^2 = 0.405\ldots$.  To see why this
is so, let ${a \over 2^m} = 0.a_1 \ldots a_m$ be the best $m$-bit
estimate of $\phi$.  Then $\phi = {a \over 2^m} + \delta$, where $0 <
|\delta| \le {1 \over 2^{m+1}}$.  Applying the inverse QFT to state
(\ref{qftphi}) yields the state
\begin{eqnarray}
  {1 \over 2^m} \sum_{x=0}^{2^m-1} \sum_{y=0}^{2^m-1} e^{-2\pi i x y
    \over 2^m} e^{2\pi i \phi y} \ket{x} & = & {1 \over 2^m}
  \sum_{x=0}^{2^m-1} \sum_{y=0}^{2^m-1} e^{-2\pi i x y \over 2^m}
  e^{2\pi i ({a \over 2^m}+\delta) y} \ket{x} \nonumber \\ & = & {1
    \over 2^m} \sum_{x=0}^{2^m-1} \sum_{y=0}^{2^m-1} e^{2\pi i (a-x) y
    \over 2^m} e^{2\pi i \delta y}\ket{x}
\label{qftstate}
\end{eqnarray}
(for clarity, we are now including the normalization factors) and the
coefficient of $\ket{a_1 \cdots a_m}$ in the above is the geometric
series
\begin{eqnarray}
  {1 \over 2^m} \sum_{y=0}^{2^m-1} (e^{2\pi i \delta})^y & = & {1
    \over 2^m} \left({1 - (e^{2\pi i \delta})^{2^m} \over 1 - e^{2\pi
      i \delta}}\right)\;.
\end{eqnarray}
Since $|\delta| \le {1 \over 2^{m+1}}$, it follows that 
$2 \pi \delta 2^m \le \pi$, and thus 
$|1 - e^{2\pi i \delta 2^m}| \ge {2 \pi \delta 2^m \over \pi / 2} = 4
\delta 2^m$.  Also, $|1 - e^{2 \pi i \delta}| \le 2 \pi \delta$.
Therefore, the probability of observing $a_1 \cdots a_m$ when
measuring the state is
\begin{equation}
  \left|{1 \over 2^m} \left({1 - (e^{2\pi i \delta})^{2^m} \over 1 -
    e^{2\pi i \delta}}\right)\right|^2 \ge \left({1 \over 2^m}
  \left({4 \delta 2^m \over 2 \pi \delta}\right)\right)^2 = {4 \over
    \pi^2}\;.
\end{equation}
Note that the above algorithm (described by networks in Figs.~\ref{fig5} 
and \ref{shorkit}) consists of $m$ controlled-$U^{2^k}$ operations, and 
$O(m^2)$ other operations.

In many contexts (such as that of the factoring algorithm of Shor), 
the above positive probability of success is sufficient to be useful; 
however, in other contexts, a higher probability of success may be 
desirable.
The success probability can be amplified to $1 - \epsilon$ for any 
$\epsilon > 0$ by inflating $m$ to $m^{\prime} = m + O(\log(1 / \epsilon))$, 
and rounding off the resulting $m^{\prime}$-bit string to its most 
significant $m$ bits.
The details of the analysis are in Appendix C.

The above approach was motivated by the method proposed by Kitaev (1995), 
which involves a sequence of repetitions for each unit $U^{2^j}$.
The estimation of $\phi$ can also be obtained by other methods, such as 
the techniques studied for optimal state estimation by Serge Massar and 
Sandu Popescu (1995), Radoslav Derka, Vladimir Buzek, and Ekert (1997), 
and the techniques studied for use in frequency standards by 
Susana Huelga, Macchiavello, Thomas Pellizzari, Ekert, Martin Plenio, 
and Ignacio Cirac (1997).
Also, it should be noted that the QFT, and its inverse, can be implemented 
in the fault tolerant ``semiclassical'' way (see Robert Griffiths and 
Chi-Sheng Niu (1996)).

\section{The Order-Finding Problem}
\label{s:shor}

In this section, we show how the scheme from the previous section can 
be applied to solve the order-finding problem, where one is given positive 
integers $a$ and $N$ which are relatively prime and such that $a < N$, 
and the goal is to find the minimum positive integer $r$ such that 
$a^r \bmod N = 1$.
There is no known classical procedure for doing this in time polynomial 
in $n$, where $n$ is the number of bits of $N$.
Shor (1994) presented a polynomial-time quantum algorithm for this problem, 
and noted that, since there is an efficient classical randomised reduction 
from the factoring problem to order-finding, there is a polynomial-time 
quantum algorithm for factoring.
Also, the quantum order-finding algorithm can be used directly to 
break the RSA cryptosystem (see Appendix A).

Let us begin by assuming that we are also supplied with a prepared
state of the form
\begin{equation}
\ket{\psi_1} = \sum_{j=0}^{r-1} e^{-2 \pi i j \over r} 
\ket{a^j \mbox{ mod } N }\;.
\end{equation}
Such a state is not at all trivial to fabricate; we shall see how
this difficulty is circumvented later.
Consider the unitary transformation $U$ that maps $\ket{x}$ to 
$\ket{a x \bmod N}$.
Note that $\ket{\psi_1}$ is an eigenvector of $U$ with eigenvalue
$e^{2\pi i ({1 \over r})}$.
Also, for any $j$, it is possible to implement a controlled-$U^{2^j}$ 
gate in terms of $O(n^2)$ elementary gates.
Thus, using the state $\ket{\psi_1}$ and the implementation of 
controlled-$U^{2^j}$ gates, we can directly apply the method of 
Sect.~\ref{s:ip} to efficiently obtain an estimator of ${1 \over r}$ 
that has $2n$-bits of precision with high probability.
This is sufficient precision to extract $r$.

The problem with the above method is that we are aware of 
no straightforward efficient method to prepare state $\ket{\psi_1}$.
Let us now suppose that we have a device for the following kind of state 
preparation.
When executed, the device produces a state of the form 
\begin{equation}
\ket{\psi_{k}} = 
\sum_{j=0}^{r-1} e^{-\frac{2\pi i k j}{r}}\ket{a^j \bmod N}\;,
\end{equation}
where $k$ is randomly chosen (according to the uniform distribution) 
from $\{1,\ldots,r\}$.
We shall first show that this is also sufficient to efficiently 
compute $r$, and then later address the issue of preparing such states.
For each $k \in \{1,\ldots,r\}$, the eigenvalue of state $\ket{\psi_{k}}$ 
is $e^{2 \pi i ({k \over r})}$, and we can again use the technique from 
Sect.~\ref{s:ip} to efficiently determine ${k \over r}$ with $2n$-bits of 
precision.
From this, we can extract the quantity ${k \over r}$ exactly by the method 
of continued fractions.
If $k$ and $r$ happen to be coprime then this yields $r$; otherwise, 
we might only obtain a divisor of $r$.
Note that, we can efficiently {\em verify} whether or not we happen to 
have obtained $r$, by checking if $a^r \bmod N = 1$.
If verification fails then the device can be used again to produce another 
$\ket{\psi_k}$.
The expected number of random trials until $k$ is coprime to $r$ 
is $O(\log\log(N)) = O(\log n)$.

In fact, the expected number of trials for the above procedure can 
be improved to a constant.
This is because, given any two independent trials which yield 
${k_1 \over r}$ and ${k_2 \over r}$, it suffices for $k_1$ and $k_2$ 
to be coprime to extract $r$ (which is then the least common denominator 
of the two quotients).
The probability that $k_1$ and $k_2$ are coprime is bounded below by 
\begin{equation}
1 - \sum_{p \mbox{\scriptsize \ prime}} 
\Pr[\,p \mbox{\ divides} \; k_1] \Pr[\,p \mbox{\ divides}\; k_2] \ge  
1 - \sum_{p \mbox{\scriptsize \ prime}} {1 / p^2} > 0.54\;.
\end{equation}

Now, returning to our actual setting, where we have no special devices 
that produce random eigenvectors, the important observation is that 
\begin{equation}
\ket{1} = \sum_{k=1}^{r} \ket{\psi_k}\;, 
\end{equation}
and $\ket{1}$ {\em is} an easy state to prepare.
Consider what happens if we use the previous quantum algorithm, but 
with state $\ket{1}$ substituted in place of a random $\ket{\psi_k}$.
In order to understand the resulting behavior, imagine if, initially, 
the control register were measured with respect to the orthonormal 
basis consisting of 
$\ket{\psi_1}$, \ldots, $\ket{\psi_{r}}$.
This would yield a uniform sampling of these $r$ eigenvectors, 
so the algorithm would behave exactly as the previous one.
Also, since this imagined measurement operation is with respect to an 
orthonormal set of eigenvectors of $U$, it commutes with all the 
controlled-$U^{2^j}$ operations, and hence will have the same effect 
if it is performed at the {\em end\/} rather than at the beginning 
of the computation.
Now, if the measurement were performed at the end of the computation 
then it would have no effect on the outcome of the measurement of the 
control register.
This implies that state $\ket{1}$ can in fact be used in place 
of a random $\ket{\psi_k}$, because the relevant information that 
the resulting algorithm yields is {\em equivalent}.
This completes the description of the algorithm for the order-finding 
problem.

It is interesting to note that the algorithm that we have described 
for the order-finding problem, which is follows Kitaev's methodology, 
results in a network (Fig.~\ref{shorkit} followed by Fig.~\ref{fig5} 
backwards) that is identical to the network for Shor's algorithm, 
although the latter algorithm was derived by an apparently different 
methodology.
The sequence of controlled-$U^{2^j}$ operations is equivalent to the 
implementation (via repeated squarings) of the modular exponentiation 
function in Shor's algorithm.
This demonstrates that Shor's algorithm, in effect, estimates the 
eigenvalue corresponding to an eigenstate of the operation $U$ that 
maps $\ket{x}$ to $\ket{ax \bmod N}$.

\section{Generating Arbitrary Interference Patterns}
\label{s:gen}

We will show in this section how to generate specific interference
patterns with arbitrary precision via some function evaluations.  We
require two registers.  The first we call the control register; it
contains the states we wish to interfere.  The second we call the
auxiliary register and it is used solely to induce relative phase
changes in the first register.

Suppose the first register contains $n$ bits.  For each $n$-bit string
$\ket{x}$ we require a unitary operator $U_x$.  All of these operators
$U_x$ should share an eigenvector $\ket{\Psi}$ which will be the state
of the auxiliary register. Suppose the eigenvalue of $\ket{\Psi}$ for
$x$ is denoted by $e^{2 \pi i\phi(x)}$.  By applying a unitary
operator to the auxiliary register conditioned upon the value of the
first register we will get the following interference pattern:

\begin{eqnarray}
  && \sum_{x=0}^{2^n-1} \ket{x}\ket{\Psi} \rightarrow \sum_{x=0}^{2^n-1}
  \ket{x}U_x(\ket{\Psi}) \\ && = \sum_{x=0}^{2^n-1} e^{2\pi i \phi(x)}
  \ket{x}\ket{\Psi} .
\end{eqnarray}

The Conditional $U_f$ gate that was described in section 2 can be
viewed in this way.  Namely, the operator $U_{f(0)}$ which maps
$\ket{y}$ to $\ket{y \oplus f(0)}$ and the operator $U_{f(1)}$ which
maps $\ket{y}$ to $\ket{y \oplus f(1)}$ have common eigenstate
$\ket{0} - \ket{1}$.  The operator $U_{f(j)}$ has eigenvalue $e^{2\pi
  i \frac{f(j)}{2}}$ for $j=0,1$.

In general, the family of unitary operators on $m$ qubits which simply
add a constant integer $k$ modulo $2^m$ share the eigenstates

\begin{equation}
  \sum_{y=0}^{2^m-1} e^{-2 \pi i \frac{ly}{2^m}} \ket{y},
\end{equation}
and kick back a phase change of $e^{2\pi i \frac{kl}{2^m}}$.

For example, suppose we wish to create the state $\ket{0} + e^{2\pi i
  \phi}\ket{1}$ where $\phi = 0.a_1 a_2 a_3 \ldots a_m$.

We could set up an auxiliary register with $m$ qubits and set it to
the state
\begin{equation}
  \sum_{y=0}^{2^m-1} e^{-2 \pi i \phi y} \ket{y}.
\end{equation}

By applying the identity operator when the control bit is $\ket{0}$ and the 'add $1$
modulo $2^m$' operator, $U_1$, when the control bit is $\ket{1}$ we see that

\[ \ket{0} \sum_{y=0}^{2^m-1} e^{-2\pi i \phi y} \ket{y} \]
gets mapped to itself and 
\[ \ket{1} \sum_{y=0}^{2^m-1} e^{-2\pi i \phi y} \ket{y} \]
goes to 

\begin{eqnarray}
&& \ket{1} \sum_{y=0}^{2^m-1} e^{-2\pi i \phi y} \ket{y+1 \bmod 2^m } \\
&& = e^{2\pi i \phi} \ket{1} 
\sum_{y=0}^{2^m-1} e^{-2\pi i \phi (y+1)} \ket{y+1 \bmod 2^m }  \\
&& = e^{2\pi i \phi} \ket{1} 
\sum_{y=0}^{2^m-1} e^{-2\pi i \phi y} \ket{y} .\\
\end{eqnarray}

An alternative is to set the $m$-bit auxiliary register to the
eigenstate
\begin{equation}
  \sum_{y=0}^{2^m-1} e^{-\frac{2\pi i}{2^m} y} \ket{y}
\end{equation}
and conditionally apply $U_{\phi}$ which adds $a = a_1a_2 \ldots a_m$ to
the auxiliary register.  Similarly, the state
\[ \ket{1} \sum_{y=0}^{2^m-1} e^{-\frac{2\pi i}{2^m} y} \ket{y} \]
goes to 
\begin{eqnarray}
  && \ket{1} \sum_{y=0}^{2^m-1} e^{-\frac{2\pi i}{2^m} y} 
  \ket{y+a \bmod 2^m } \\ && = e^{2\pi i \phi} \ket{1} \sum_{y=0}^{2^m-1}
  e^{-\frac{2\pi i}{2^m} (y+ a)} \ket{y+ a \bmod 2^m} \\ && =
  e^{2\pi i \phi} \ket{1} \sum_{y=0}^{2^m-1} e^{-\frac{2\pi i}{2^m} y} 
\ket{y} .
\end{eqnarray}

Similarly, if $\phi = ab/2^m$ for some integers $a$ and $b$, we could
also obtain the same phase ``kick-back'' by starting with state

\begin{equation}
\sum_{y=0}^{2^m-1} e^{-2\pi i \frac{a}{2^m} y} \ket{y}
\end{equation}

and conditionally adding $b$ to the second register.

The method using eigenstate
\begin{equation}
  \sum_{y=0}^{2^m-1} e^{-\frac{2\pi i}{2^m} y} \ket{y}
\end{equation}
has the advantage that we can use the same eigenstate in the auxiliary
register for any $\phi$.  So in the case of an $n$-qubit control
register where we want phase change $e^{2\pi i \phi(x)}$ for state
$\ket{x}$ and if we have a reversible network for adding $\phi(x)$ to
the auxiliary register when we have $\ket{x}$ in the first register,
we can use it on a superposition of control inputs to produce the
desired phase ``kick-back'' $e^{2\pi i \phi(x)}$ in front of $\ket{x}$.
Which functions $\phi(x)$ will produce a useful result, and how to
compute them depends on the problems we seek to solve.
 
\section{Conclusions}
\label{s:conc}

Various quantum algorithms, which may appear different, exhibit 
remarkably similar structures when they are cast within the paradigm 
of multi-particle interferometry.
They start with a Fourier transform to prepare superpositions of 
classically different inputs, followed by function evaluations 
(i.e. $f$-controlled unitary transformations) which induce 
interference patterns (phase shifts), and are followed by another 
Fourier transform that brings together different computational paths 
with different phases.
The last Fourier transform is essential to guarantee the 
interference of different paths.

We believe that the paradigm of estimating (or determining exactly) 
the eigenvalues of operators on eigenstates gives helpful 
insight into the nature of quantum algorithms and may prove useful 
in constructing new and improving existing algorithms.
Other problems whose algorithms can be deconstructed in a similar 
manner are: Simon's algorithm (1993), Shor's discrete logarithm 
algorithm (1994), Boneh and Lipton's algorithm (1995), 
and Kitaev's more general algorithm for the Abelian Stabiliser 
Problem (1995), which first highlighted this approach.

We have also shown that the evaluation of classical functions on quantum 
superpositions can generate arbitrary interference patterns with any 
prescribed precision, and have provided an explicit example of a universal
construction which can accomplish this task.

\begin{acknowledgments}
We wish to thank David Deutsch, David DiVincenzo, Ignacio Cirac
and Peter H\o yer for useful discussions and comments.
	
  This work was supported in part by the European TMR Research Network
  ERP-4061PL95-1412, CESG, Hewlett-Packard,   The Royal Society London, 
  the U.S. National Science Foundation under Grant No. PHY94-07194, 
  and Canada's NSERC.
  Part of this work was completed during the 1997 Elsag-Bailey -- I.S.I. 
  Foundation research meeting on quantum computation.
\end{acknowledgments}

\begin{appendix}
\section{Cracking RSA}
What we seek is a way to compute $P$ modulo $N$ given $P^e,e$, and
$N$, that is, a method of finding $e$th roots in
the multiplicative group of integers modulo $N$ (this group is 
often denoted by
$\boldmath{Z_N}^{\star}$ and contains the integers coprime to $N$).  
It is still an open question whether a
solution to this problem necessarily gives us a polynomial time randomised
algorithm for factoring.  However factoring does give a polynomial
time algorithm for finding $e$th roots for any $e$ relatively prime to
$\phi(N)$ and thus for cracking RSA.  Knowing the prime factorisation
of $N$, say $\prod p_1^{a_1} p_2^{a_2} \ldots p_k^{a_k}$, we can
easily compute $\phi(N) = N \prod_{i=1}^{n}(1-\frac{1}{p_i})$.  Then
we can compute $d$ such that $ed \equiv 1$ mod $\phi(N)$, which
implies $P^{ed} \equiv P$ modulo $N$.

However, to crack a particular instance of RSA, it suffices to find an
integer $d$ such that $ed \equiv 1$ modulo ord($P$), that is $ed =
\mbox{ord}(P)k+1$ for some integer $k$.  We would then have $C^d
\equiv P^{ed} \equiv P^{\mbox{ord}(P)k+1} \equiv P$ modulo $N$.

Since $e$ is relatively prime to $\phi(N)$ it is easy to see that
$\mbox{ord}(P) = \mbox{ord}(P^e) = \mbox{ord}(C)$.  So given $C =
P^e$, we can compute $\mbox{ord}(P)$ using Shor's algorithm 
and then compute $d$
satisfying $de \equiv 1$ modulo $\mbox{ord}(P)$ using the extended
Euclidean algorithm.  Thus, we do not need several repetitions of
Shor's algorithm to find the order of $a$ for various random $a$;
we just find the order of $C$ and solve for $P$ regardless of
whether or not this permits us to factor $N$.

\section{Concatenated Interference}

\begin{figure}[!hb]
  \vspace*{0.2cm} \centerline{\psfig{width=12cm,file=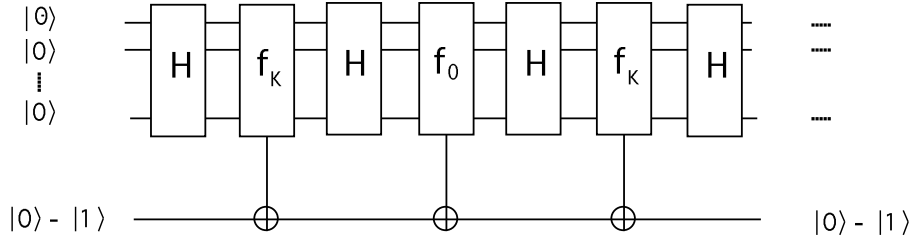}}
  \vspace*{5mm}
\caption{Network representation of Grover's algorithms. By repeating
  the basic sequence $2^{n/2}$ times, value $k$ is obtained at the
  output with probability greater than $0.5$.}
\label{grover}
\end{figure}

The generic sequence: a Hadamard/Fourier transform, followed by an
$f$-controlled-$U$, followed by another Hadamard/Fourier transform can
be repeated several times. This can be illustrated, for example, with
Grover's data base search algorithm (1996). Suppose we are given (as
an oracle) a function $f_k$ which maps $\{0,1\}^n$ to $\{0,1\}$ such
that $f_k(x)=\delta_{xk}$ for some $k$.  Our task is to find $k$. Thus
in a set of numbers from $0$ to $2^{n}-1$ one element has been
``tagged'' and by evaluating $f_k$ we have to find which one. To find
$k$ with probability of $50\%$ any classical algorithm, be it
deterministic or randomised, will need to evaluate $f_k$ a minimum of
$2^{n-1}$ times.  In contrast, a quantum algorithm needs only
$O(2^{n/2})$ evaluations.  Grover's algorithm can be best presented as
a network shown in Fig.~\ref{grover}.
 
\section{Amplifying success probability when estimating phases}

Let $\phi$ be a real number satisfying $0 \leq \phi < 1$ which is not
a fraction of $2^m$, and let 
$\frac{a}{2^m} = 0.a_1 a_2 \ldots a_m$ be the closest $m$-bit
approximation to $\phi$ so that $\phi = \frac{q}{2^m} + \delta$
where $0 < |\delta | \leq \frac{1}{2^{m+1}}$.
For such a $\phi$, we have already shown that 
applying the inverse of the QFT to
(\ref{qftphi}) and then measuring yields the state $\ket{a}$
with probability at least $4 / \pi^2 = 0.405\ldots$.  

WLOG assume $0 < \delta \leq \frac{1}{2^m+1}$.
For $t$ satisfying $-2^{m-1} \leq t < 2^{m-1}$ let $\alpha_{t}$
denote the amplitude of $\ket{a-t \bmod 2^m}$.
It follows from
(\ref{qftstate}) that
\begin{equation}
\alpha_t = {1 \over 2^m} 
\left({1 - (e^{2\pi i (\delta + \frac{t}{2^m})})^{2^m} 
\over 1 - e^{2\pi i (\delta + \frac{t}{2^m})}}\right)\;.
\end{equation}
Since 
\begin{equation}
\left| 1 - e^{2\pi i (\delta + \frac{t}{2^m})} \right| \leq 
{ 2\pi (\delta +  \frac{t}{2^m}) \over \pi/2} = 4 (\delta + \frac{t}{2^m})
\end{equation}
then 
\begin{equation}
 \left| \alpha_{t} \right| 
\leq \left| {2 \over 2^m 4 (\delta + \frac{t}{2^m})} \right| 
\leq {1 \over 2^{m+1} (\delta + \frac{t}{2^m})}.
\end{equation}

The probability of getting an error greater than $\frac{k}{2^m}$
is
\begin{eqnarray}
&& \sum_{k \leq t < 2^{m-1}} \left| \alpha_t \right|^2
+ \sum_{-2^{m-1} \leq t < -k} \left| \alpha_t \right|^2
\\
&& 
\leq \sum_{t= k}^{2^{m-1}-1} {1 \over 4(t+2^m\delta)^2}
+ \sum_{t= -2^{m-1}}^{-(k+1)} {1 \over 4(t+2^m\delta)^2}
\\
&& 
\leq \sum_{t= k}^{2^{m-1}-1} {1 \over 4t^2}
+ \sum_{t= k+1}^{2^{m-1}} {1 \over 4(t-\frac{1}{2})^2}
\\
&& 
\leq \sum_{t= 2k}^{2^{m}-1} {1 \over 4(\frac{t}{2})^2}
\\
&& 
< \int_{2k-1}^{2^m-1} {1 \over t^2}
\\
&& 
< {1 \over 2k-1}.
\\
\end{eqnarray}

So, for example, if we wish to have
an estimate that is within $1/2^{n+1}$ of the value
$\phi$ with probability at least $1-\epsilon$ it suffices
to use this technique with 
$m = n+\lceil \log_2{(\frac{1}{2\epsilon} + \frac{1}{2}) } \rceil$ 
bits.

\end{appendix}

\end{document}